\documentstyle[prb,multicol,aps,epsfig]{revtex}

\begin{document}

\title{Electronic structure and light-induced conductivity
in a transparent refractory oxide}

\author{J.~E. Medvedeva\(^{1,*}\), A.~J. Freeman\(^{1}\),
M.~I. Bertoni\(^{2}\) and T.~O. Mason\(^{2}\)
}
\address{\(^1\)Department of Physics and Astronomy and
\, \(^2\)Department of Materials Science and Engineering, 
Northwestern University, Evanston, Illinois 60208-3112
}

%\date{\today}
\maketitle

\begin{abstract}
Combined first-principles and experimental investigations reveal 
the underlying mechanism responsible for a drastic change of 
the conductivity (by 10 orders of magnitude) following hydrogen annealing 
and UV-irradiation in a transparent oxide, 12CaO$\cdot$7Al$_2$O$_3$,
found by Hayashi {\it et al}. 
The charge transport associated with photo-excitation 
of an electron from H, occurs by electron hopping.
We identify the atoms participating in the hops, determine the exact paths 
for the carrier migration, estimate the temperature behavior
of the hopping transport and predict a way 
to enhance the conductivity by specific doping.
\end{abstract}


\begin{multicols}{2}

The rare combination of transparency and conductivity
has a wide range of technological applications \cite{TCO}.
Until recently, optimum optical transmission and useful electrical
conductivity ($>$10$^3$~S~cm$^{-1}$) was attained by the introduction 
of selected dopants into a wide-bandgap oxide. 
Hence the excitement surrounding the report that 
Hayashi {\it et al} found a new way to convert a transparent 
oxide into a persistent electronic conductor \cite{Hayashi2002} 
with the potential for less expensive and more flexible optoelectronic 
devices (including flat panel displays, solar cells, invisible circuits)
as well as three-dimensional holographic memories.
Despite careful experimental studies, no definitive understanding 
has been reached on the underlying mechanism responsible 
for this new dramatic effect.
In this combined theoretical and experimental investigation,
we address the nature of this novel phenomenon which 
is crucial for further progress and make predictions 
on how to improve the conductivity toward the level 
where it becomes useful for practical applications.

We have reproduced in bulk form the single crystal and thin film results 
of Hayashi {\it et al} by solid state reaction of 
high purity oxides \cite{Hosono87} (99.99+\%),
followed by hydrogen treatment (5\%H$_2$/95\%N$_2$) of sintered pellets 
at 1,300~$^{\circ}$C for 2~h, and by quenching to room temperature. 
UV-treatment was performed under a mercury arc lamp (40~min at 20~mW~cm$^{-2}$ 
and a wavelength of 220~nm). This turned a surface layer of $\sim$16~$\mu m$ 
thickness from white to green, with a conductivity changed from $10^{-10}$
to 0.6~S~cm$^{-1}$ (by four-point method).  
We also irradiated loose H-treated powders, which also turned from white 
to a uniform green color, and measured their conductivity by a 
``powder-solution-composite'' method \cite{Mason} to be 0.41~S~cm$^{-1}$.  
Both values agree with those in Ref. \onlinecite{Hayashi2002}
and provide confidence in our experimental confirmation of the predictions
reported below.

From the theoretical side, we demonstrate with density functional 
calculations that the conductivity appears to result from
electrons excited off the hydrogen ions into conduction bands
formed from Ca 3d states. The changes found in the electronic
properties caused by the UV-irradiation of H-doped mayenite allow us 
to conclude that the observed absorption peak at 0.4 eV has a fundamentally 
different nature than the F$^+$ center energy level originally assumed 
in Ref. \onlinecite{Hayashi2002} and followed
in recent theoretical work \cite{Sushko03}. 
Our calculated charge density shows clearly that electron trapping on 
a vacancy (i.e., formation of the F$^+$ center) does not occur 
in the case of UV-irradiated H-doped mayenite. Instead,  
we identify the atoms participating in hopping transport;
estimates of the characteristic parameters that determine the behavior
of the hopping conductivity at low (Efros-Shklovskii regime)
and high (Mott) temperatures agree reasonably well with experiment.
Finally, we found a strong dependence of the electronic
transport on the particular hopping centers and their spatial
arrangement that allows the prediction of rates of temperature
treatment and dopants which will enhance the conductivity.

\begin{figure}
\includegraphics[width=4.2cm]{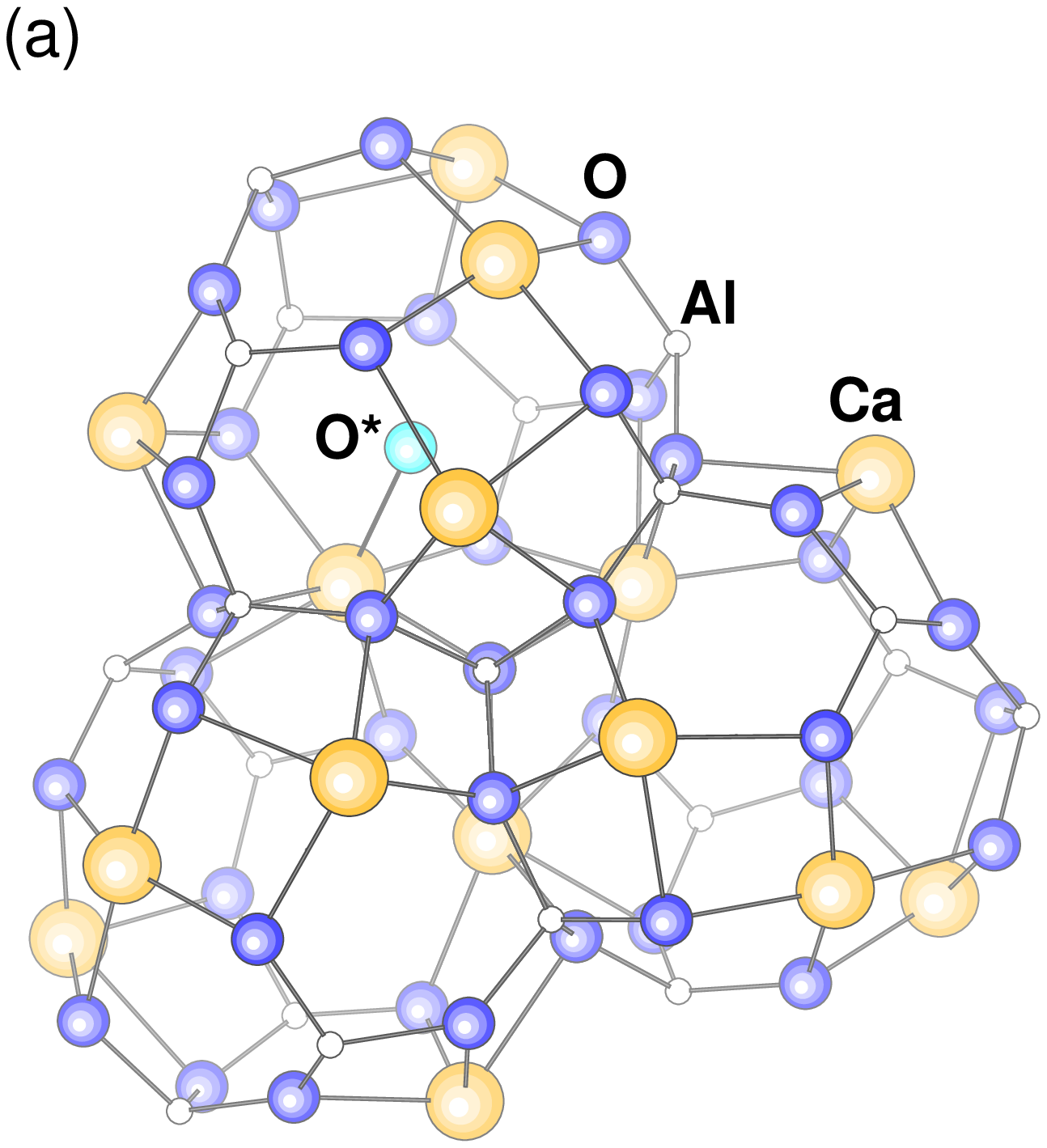}
\includegraphics[width=4.0cm]{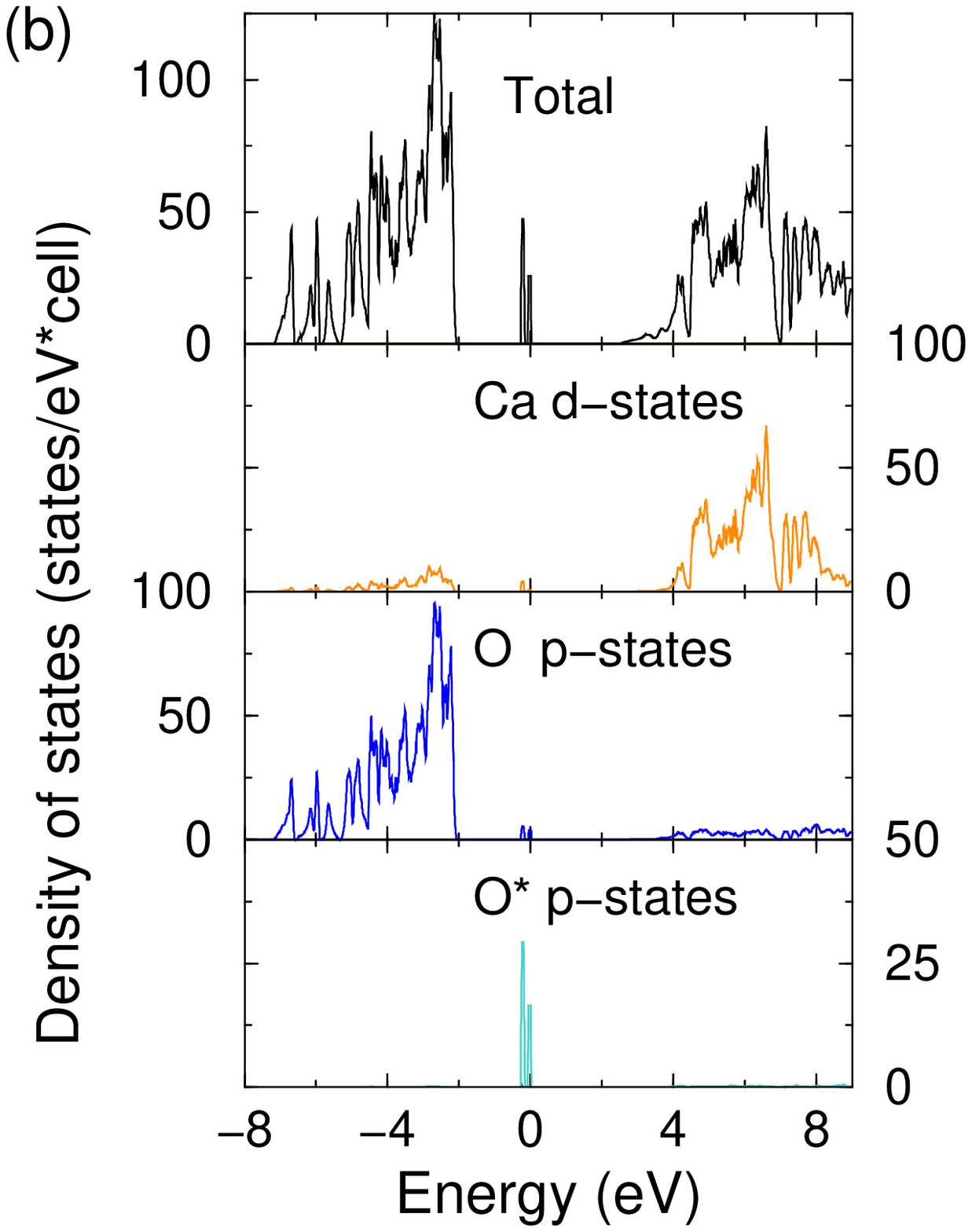}
\caption{
(a) Three of the 12 cages constituting a unit cell of mayenite. 
An O$^{2-}$ anion (abbreviated as O$^*$) is located inside one of six cages
so that the distances between O$^*$ and its six neighbor Ca atoms  
vary from 2.36 to 3.95 \AA.
(b) Total and partial DOS of 12CaO$\cdot$7Al$_2$O$_3$; E$_F$ is at 0 eV.
The Al 3d states form the band centered at $\sim$28 eV.
}
\label{fig1}
\end{figure}

The crystal structure of 12CaO$\cdot$7Al$_2$O$_3$
has the space group $I\overline43d$, with two formula units 
in the unit cell and a lattice parameter of 11.98 \AA \, 
(Ref. \onlinecite{struct,Imlach71}).
This framework includes 64 of the oxygen atoms; the remaining two oxygen 
ions (abbreviated as O$^*$ hereafter) which provide charge neutrality 
\cite{Imlach71} are located inside the cages, Fig. \ref{fig1}(a). 
These two O$^*$ are distributed 
between 24 sites that produce a structural disorder. 
From full-potential linearized augmented plane-wave \cite{flapw} (FLAPW) 
total energy calculations of seven structures 
with different distances between these two oxygen ions 
(ranging from 5.1 \AA \, to 10.4 \AA), 
we found that the O$^*$s tend to maximize the distance between 
them by forming a bcc lattice with the crystallographic basis oriented
randomly with respect to that of the whole crystal.
After hydrogen annealing, changes in the crystal structure 
are not observed \cite{Dravid}.
The incorporation of H$^-$ ions corresponds to the following chemical reaction: 
O$^{2-}$ (cage) + H$_2$ (atm.) $\rightarrow$ OH$^-$ (cage) + H$^-$ 
(in another cage), so the unit cell contains two cages occupied by O$^*$H, 
another two by hydrogen ions (abbreviated as H$^*$ hereafter) and the 
remaining 8 cages are empty \cite{dmol}. 
From the electronic structure calculations, 
we found that the structure with two H$^-$ ions in cages and 
released O$_2$ gas (cf., Ref. \onlinecite{Hayashi2002})
leads to an insulating state after photo-activation.

\begin{figure}
\includegraphics[width=8.3cm]{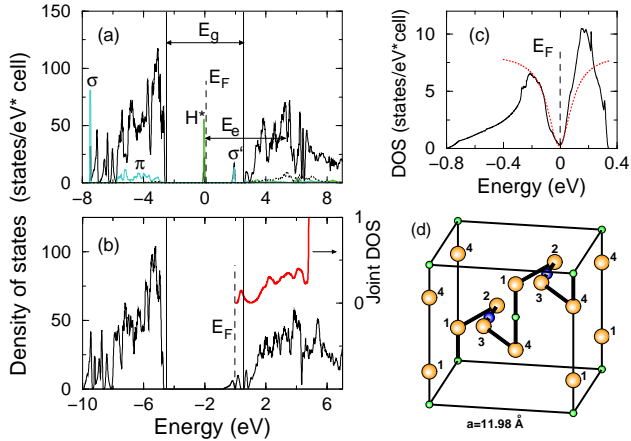}
\caption{
(a) Total DOS of H-bearing mayenite. The band gap, E$_g$, is between 
O 2p and Ca 3d states. The DOS of the impurity bands 
arising from the OH$^-$ complex and H$^*$ is enlarged by a factor of five. 
The photo-activation of the H$^*$ corresponds to the transition 
from H$^*$ 1s states to the 3d states of its closest Ca 
(dashed line), with energy E$_e$.
(b) Total DOS of UV-irradiated H-doped mayenite 
and the joint DOS (thick line) calculated using the optical selection 
rules and only nearest neighbor transitions which have higher 
probability.
(c) Enlarged total DOS of UV-irradiated H-doped mayenite
and a theoretical fit to the DOS near E$_F$ 
(dashed line), see text.
(d) The optimal path for electron migration throughout
the crystal. The cube represents a full unit cell 
consisting of 12 cages (118 atoms); only atoms which give the dominant 
contribution to the band crossing E$_F$ are shown -- Ca (largest spheres), 
O$^*$ (between Ca(2) and Ca(3)) and H$^*$ (smallest spheres).
}
\label{fig2}
\end{figure}

The calculated \cite{lmto} electronic density of states (DOS) before H-doping
is shown in Fig. \ref{fig1}(b): oxygen 2p states form the top of the valence
band between $-$6 eV and $-$2 eV (the Fermi level is taken as zero), 
while the bottom of the conduction band is formed from Ca 3d states. 
Located inside one of the six cages, O$^*$ 
gives a fully occupied impurity peak below $E_F$. 
The calculated band gap, 4.8 eV, underestimates the experimental 
absorption edge \cite{Hayashi2002} by 4 \%. 
The incorporation of H ions into the cages of insulating 
mayenite results in the appearance of new bands 
(Fig. \ref{fig2}(a)): (i) since H shares its electron with O$^*$
(and forms an OH$^-$ complex), 
filled $\sigma$ and non-bonding $\pi$ bands and unoccupied 
$\sigma '$ band are introduced; and (ii) the 1s states of H$^*$ 
form the fully occupied impurity band below E$_F$.
After the H-doping, the energy of the transitions from  
O 2p states to Ca 3d states is decreased by 0.7 eV (from 5.9 eV to 5.2 eV), 
in agreement with experiment \cite{Hayashi2002}.
The maximum efficiency of the photo-activation of the H$^*$, 
which corresponds to transitions from H$^*$ 1s states 
to the 3d states of its closest Ca neighbors, 
occurs at about the same photon energy (Fig. \ref{fig2}(a)). 
These results are in contrast with the conclusion of 
Ref. \onlinecite{Hayashi2002} that hydrogen controls the absorption edge.

For UV-irradiated H-bearing mayenite, the electron excitation
from H$^*$ is calculated with the model system where the electron 
is transferred equally to the d states of six Ca neighbors \cite{tb}. 
The self-consistent DOS shows the formation of a new band 
(Fig. \ref{fig2}(b))
with the same total number of states, $m$, as in the band gap 
of the system before UV irradiation. However, since $m$ is now two times 
larger than the number of available electrons, $E_F$ passes through 
the middle of the band and the system becomes conducting.
The calculated positions of the characteristic absorption bands
(0.38 eV and 3.5 eV, Fig. \ref{fig2}(b)) 
are in good agreement with experiment (0.4 eV and 2.8 eV), 
as is the intense charge transfer transition from oxygen 2p states 
to Ca 3d states which occurs at $\sim$4.8 eV.

\begin{figure}
\includegraphics[width=8.5cm]{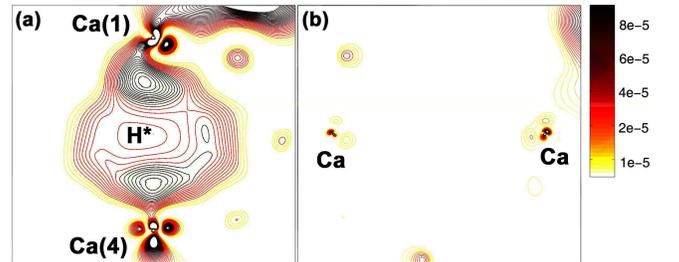}
\caption{
Contour map of the charge density distribution within a slice passing through
the center of (a) a cage with H$^*$ and (b) an empty one (vacancy).}
\label{charge}
\end{figure}

As shown in Fig. \ref{fig2}(c), the non-zero DOS at E$_F$, $g(E_F)$, 
is determined mainly by H$^*$, O$^*$ and Ca states -- atoms 
that are spatially well separated from each other (Fig. \ref{fig2}(d)),
which points to the hopping nature of the induced conductivity.
It is striking that the arrangement of the atoms
corresponds to the shortest electron hopping path:
only the closest Ca neighbors of O$^*$, Ca(2) and Ca(3)
(at distances of 3.3 and 2.4 \AA, respectively),
and the closest Ca neighbors of H$^*$, Ca(1) and Ca(4) 
(both at 2.8 \AA), 
give significant contributions to $g(E_F)$ (Table \ref{table}).
The longest segment on the hopping path corresponds to 
the distance between Ca(1), Ca(2) and O$^*$, 
which explains the higher probability to find the electron on Ca(1) and Ca(2). 
Finally, the charge density distribution calculated in the energy 
window of 25 meV below E$_F$, clearly shows the connected 
electron density maxima along the hopping path (cf., Fig. \ref{charge}(a)). 
The charge density in the empty cages is insignificant
(cf., Fig. \ref{charge}(b)), demonstrating that
trapping of the electron on the vacancy does not occur \cite{Ocalc}.

Hopping transport in systems with localized electronic states
has been studied analytically, numerically and experimentally 
\cite{Mott,ES84,hopping}.
At high temperatures (T) the hopping occurs between nearest neighbors 
and has an activated nature. At lower T, the activation 
energy for hops between localized centers can be reduced by enlarging 
the hopping distance -- which leads to the variable-range hopping 
described by Mott \cite{Mott}. At yet lower T, 
the electron-electron repulsive interaction between charge carriers 
may result in the formation of a Coulomb gap centered at E$_F$. 
The DOS at E$_F$ may not turn exactly to zero, corresponding
to a `soft' Coulomb gap \cite{ES84}:
this residual DOS at E$_F$ is caused by correlated simultaneous hops 
of several electrons (multi-electron hopping) which plays a dominant 
role \cite{Pollak79} at extremely low T.

Similar to the above qualitative picture,
we predict the behavior of the conductivity at finite T
based on the calculated DOS in the vicinity of E$_F$
for the irradiated H-doped mayenite. Clearly, this exhibits in 
Fig. \ref{fig2}(c) a soft Coulomb gap near E$_F$ and, similar 
to Fig. 10.1 in Ref. \onlinecite{ES84}, has two peaks. 
These results conform to the Efros-Shklovskii (ES) 
behavior \cite{ES84} of the conductivity $\sigma$ at low T: 
\begin{equation}
\sigma = \sigma_0 exp [-(T_{ES}/T)^{1/2}], \hspace{1cm} T_{ES}=\beta_{ES}e^2/k_Bk\xi,
\end{equation}
where $\xi$ is the localization length, $e$ -- the elementary charge,
$k_B$ -- the Boltzmann constant, and $k$=$\varepsilon_0 \varepsilon_r$ 
is the dielectric constant ($\varepsilon_0$ -- the permittivity constant
and $\varepsilon_r$ -- the relative dielectric constant).
When T increases, the Coulomb gap fills.
According to Ref. \onlinecite{Sarvestani95}, a substantial DOS should be observed
already at T$\sim$0.05$\Delta \epsilon$, where $\Delta \epsilon$ 
is the gap width. At this T, the changeover to the 
Mott behavior \cite{Mott} should occur:
\begin{equation}
\sigma = \sigma_0 exp [-(T_M/T)^{1/4}], \hspace{1cm} T_M=\beta_M/k_Bg_0\xi^3.
\end{equation}
The double peak structure completely disappears at $T$$\sim$ 0.4$\Delta \epsilon$. 
Using $\Delta \epsilon$$\simeq$0.3 eV (Fig. \ref{fig2}(c)),
we obtain the crossover temperature between Mott and ES behavior to be
T$_c$$\simeq$170~K. Indeed, for H-doped UV-irradiated mayenite, 
the $T^{-1/4}$ dependence of $log (\sigma)$ was found experimentally  
\cite{Hayashi2002} for 50$<$T$<$300~K.
Further, the calculated DOS near E$_F$ fits well (Fig. \ref{fig2}(c)) 
with the theoretical one of the form
$g(E)$=$\alpha E_c^2E^2/(E_c^2+E^2)$, where $\alpha$=$(3/\pi)(k^3/e^6)$.
In the high energy limit, the DOS has the constant value
of $g_0$$\equiv$$\alpha E_c^2$ that corresponds to Mott behavior. 
In the opposite limit, it approaches
a parabolic dependence (ES) which has only one physical parameter -- 
the refractive index, n. We found n$\simeq$2.1, 
while the experiment \cite{Zhmoidin83} gives n=1.6.
From the fit, we obtain g$_0$$\simeq$8 
states/eV$\cdot$cell and $\alpha$$\simeq$800 which allows us to 
estimate the characteristic T:
using $\beta_M$=7.6 (Ref. \onlinecite{Mott93}), $\beta_{ES}$=2.8 
(Ref. \onlinecite{ES84}) and the localization length $\xi$=1.5 \AA \,
(which is comparable to the ionic radii of the atoms involved in the hops), 
we obtain T$_M$$\simeq$5.6$\times$10$^6$~K which agrees fairly well with 
the experimental value \cite{Hayashi2002}, T$_M$=2$\times$10$^6$~K,
and predict that T$_{ES}$$\simeq$2.7$\times$10$^4$~K. 
Thus, our band structure calculations
give reasonable order of magnitude estimates for the 
characteristic T of the hopping transport in mayenite.
Of course, more precise determinations require temperature 
and time-dependent calculations.

\begin{figure}
\centerline{
\includegraphics[width=5.3cm]{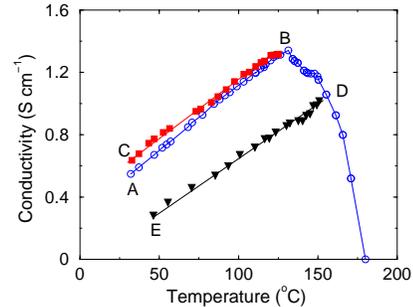}
}
\caption{
Measured $\sigma$-T characteristics 
of the UV-irradiated H-doped 12CaO$\cdot$7Al$_2$O$_3$ bulk samples
of 46\% density. 
Due to the realization of the optimal arrangement of the hopping centers 
during the heat treatment (A$\rightarrow$B), the conductivity is enhanced 
after cooling the sample (B$\rightarrow$C). 
Subsequent T cycling (B$\leftrightarrow$C) does not result in 
any changes of $\sigma$, i.e., the system is reversible. 
Cooling from any point above the conductivity maximum, 
as in path (D$\rightarrow$E), shows reduced $\sigma$ 
unchanged by UV irradiation at room T (E$\leftrightarrow$D), 
as expected from the release of H. 
}
\label{exp}
\end{figure}

Since H$^*$ may also occupy cages located farther away from 
the one with an O$^*$H, we investigated this arrangement and 
found it to be less favorable energetically -- by 96 meV. 
However, under the rapid cooling of the sample annealed at 
1,300 $^{\circ}$C some H$^*$ can become ``frozen'' 
into these positions. After UV irradiation, the corresponding 
hopping path involves a very long segment between two Ca atoms 
(now increased from 3.7 to 6.5 \AA), so that the DOS at E$_F$ 
becomes zero and the hopping has an activated behavior.
If now T is increased to a level sufficient for 
hydrogen migration, these H$^*$ will be able to diffuse into the 
energetically favorable positions which facilitates 
the hopping and leads to an increase of the conductivity.
We confirmed these predictions experimentally, as shown in Fig. \ref{exp}. 
Moreover, our samples showed non-reversible behavior once T
is increased above T$_m$, the conductivity maximum: the decay of the 
conductivity is associated with hydrogen release from the sample which
breaks up the hopping path.

If one focusses on the strong dependence of the photo-induced conductivity
on the particular atoms participating in the hopping and their
spatial arrangement, one is able to predict
the possibility of varying the conductivity by proper doping.
For example, we expect that Mg substitution
can lead to a decrease in the conductivity once Mg substitutes
one of the Ca atoms involved in the hopping, since its 3d states 
will lie much higher in energy than that of Ca, while Sr substitution
may not result in any significant changes in the conductivity.
Instead, one way to possibly improve the electronic transport of the material  
consists in increasing the concentration of hopping centers.
For this, the poorly bonded O$^{2-}$ ion inside a cage of mayenite, 
is easily replaced by 2F$^-$ or 2Cl$^-$ ions that will allow some
enhancement of the conductivity since the number of hopping centers 
and the number of carriers is now doubled.
Recent observations agree with our model and its predictions: proton implantation 
was found to increas the conductivity as compared with 
thermal H-treatment \cite{Miyakawa2003}. Our additional calculations 
show that along with the increased concentration of the encaged 
electrons, the proton implantation 
will result in the appearance of new unoccupied states in the band gap 
making H$^+$ one of the hopping centers that creates a conductivity 
channel and so enhances the transport.

Finally, the encouraging findings obtained in this first {\it ab-initio}
study of hopping transport suggest that important results may also be
obtained when a similar approach is applied to such systems as
doped semiconductors, manganite perovskites, 
carbon nanotubes, etc \cite{Pollak02}.

We thank V.P.~Dravid for sharing his unpublished electron microscopy results.
Work supported by the DOE (grant N DE-FG02-88ER45372)
and the NSF through its MRSEC program at the Northwestern
Materials Research Center. Computational resources have been provided by 
the DOE supported NERSC.
%National Energy Research Scientific Computing Center.

$^*$ Corresponding author. 

Email address: j-medvedeva@northwestern.edu

\begin{table}
\caption{The relative contributions per atom to the DOS at E$_F$.
Similar percentages are obtained from the integrated DOS 
in the energy range of E$_F\pm$0.025 eV.
}
\begin{tabular}{lclclc}
Atoms &  \% & Atoms & \%  & Atom & \% \\ \hline
Ca(1) & 7.6 & Ca(4) & 4.2 & rest Ca & $\sim$1.0 \\
Ca(2) & 8.4 & H$^*$ & 6.9 & rest O  & $\sim$0.7 \\
O$^*$ & 8.3 & Centers of &  & Al      & \, 0.1  \\
Ca(3) & 3.8 & empty cages & 0.3 &     &      \\
\end{tabular}
\label{table}
\end{table}

\end{multicols}

\begin{thebibliography}{99}
\bibitem{lmto}
Solutions of the effective one-electron
equations are obtained by means of the linear muffin-tin orbital 
method in the atomic sphere approximation \cite{Andersen}.
The band structure calculations were performed for the cell of 
12CaO$\cdot$7Al$_2$O$_3$ with one formula unit, i.e. a total 
of 59 atoms per cell which combine into six cages. 
In addition, 85 empty spheres were included to fill out the open
space of the system.
\bibitem{tb}
For the investigations of the UV-irradiated mayenite, we used 
the virtual crystal approximation which accounts properly for the 
band-filling effects.
\end{thebibliography}
\end{document}